\documentclass{aastex631}
\usepackage{amsmath}
\usepackage{colonequals}
\shorttitle{Eruptivity of Flaring Active Regions }
\shortauthors{Muhamad et al.}
\graphicspath{{./}{figures/}}

\begin{document}

\title{Eruptivity of Flaring Active Regions Based on \\
Electric Current Neutralization and Torus Instability Analysis
}

\correspondingauthor{Johan Muhamad}
\email{johan.muhamad@brin.go.id}

\author[0000-0001-7214-3754]{Johan Muhamad}
\affiliation{Research Center for Space, National Research and Innovation Agency (BRIN) \\
KST Samaun Samadikun, Jalan Sangkuriang 1-9, Coblong,  \\
Bandung, West Java 40135, Indonesia}

\author{Kanya Kusano}
\affiliation{Institute for Space-Earth Environmental Research (ISEE), Nagoya University \\
Furo-cho, Chikusa-ku,  \\
Nagoya 464-8601 Japan}



\begin{abstract}

Solar flares are frequently accompanied by coronal mass ejections (CMEs) that release significant amount of energetic plasma into interplanetary space, potentially causing geomagnetic disturbances on Earth. However, many solar flares have no association with CMEs. The relationship between solar flare and CME occurrences remains unclear. Therefore, it is valuable to distinguish between active regions that potentially produce flares and CMEs and those that do not. It is believed that the eruptivity of a flare can be characterized by the properties of the active region from which it originates. In this study, we analyzed selected active regions that produced solar flares with and without CMEs during solar cycle 24. We carefully calculated the electric current neutralization of each active region by selecting relevant magnetic fluxes based on their connectivities using nonlinear force-free field models. Additionally, we analyzed their stabilities against the torus instability by estimating the proxies of critical heights of the active regions. We found that several non-eruptive active regions, which lacked clear signatures of neutral electric currents, exhibited a more apparent relationship with high critical heights of torus instability. Furthermore, we introduced a new non-dimensional parameter that incorporates current neutralization and critical height. We found that analysing ARs based on this new parameter can better discriminate eruptive and non-eruptive flare events compared to analysis that relied solely on current neutralization or torus instability. This indicates that torus instability analysis is necessary to complement electric current neutralization in characterizing the eruptivity of solar flares. 

\end{abstract}

\keywords{}
\keywords{instabilities  --- Sun: corona --- Sun: coronal mass ejections (CMEs)--- Sun: flares --- Sun: photosphere}


\section{Introduction} \label{sec:intro}

Eruptive flares, flares that are accompanied by coronal mass ejections (CMEs), pose significant potential to induce geomagnetic storms on Earth. Consequently, characterizing the distinguishing properties of eruptive versus confined flares is important. Furthermore, recent studies have focused on identifying robust parameters for eruptivity prediction. Many previous works explored photospheric active region properties derived from magnetogram data to identify reliable eruptivity parameters \citep{nindos_2004, kazachenko_2012, sun_2015, liu_2016, sarkar_2018, lin_2020,lin_2021, li_2024}. On the other hand, others focus on the role of decay index of the background magnetic fields, particularly which lying above ARs, on the eruptivity of solar flares \citep{guo_2010, jing_2015, nindos_2012}.

\citet{harra_2016} found that there is no clear association between flare eruptivity and several physical properties of AR such as area, duration, flare ribbon area, or non-thermal energy. They found only the occurrence of coronal dimming in lines characteristic of the quiet-Sun corona that well associate to the eruptive flares. By using machine learning approach applied to large dataset of many ARs’ properties, \citet{bobra_2016} found that no single decisive parameter to discriminate between eruptive and non-eruptive flares, rather only by combining at least six parameters, a relatively effective CME prediction scheme can be made. They found that these six parameters are all intensive, which comprised of population means or related to specific source region. On the other hand, extensive parameters, which are related to sum or total quantities over an AR scale, are less important to predict a CME. For further reading of the characteristics of CME-productive solar active regions, please see \citet{kontogiannis_2023}.

Several new parameters, combining multiple well-known ones, have been proposed to differentiate between eruptive and confined flares. Many of these parameters are defined as the ratio of certain proxies of magnetic non-potentiality to the overlying magnetic field strength in an active region \citep{wangzhang_2007, thalmann_2015}. These ratios are reasonable, as flare eruptiveness is widely believed to be determined by a combination of strong energy release from below and weak strapping force from above. Several proposed schemes include the ratio of magnetic field strength to its background field \citep{toriumi_2017}, the ratio of flare ribbon area to the total AR area \citep{toriumi_2017,hinterreiter_2018, kazachenko_2022}, the ratio of highly twisted magnetic flux to its overlying field \citep{muhamad_2018,lin_2020,lin_2021}, and the ratio of current-carrying flux to total helicity \citep{thalmann_2019}. Recently, \citet{li_2022}, based on their analysis of 106 solar flares, suggested that the ratio of the mean characteristic twist parameter ($\alpha$) to the total unsigned flux can be used to distinguish between eruptive and confined flares. Although analyses of all these parameters indicate a tendency for the ratio of non-potentiality to the surrounding field to be crucial for determining flare eruptiveness, none of these parameters are canonical, as some eruptive and non-eruptive flare events still overlap around the defined thresholds for each parameter.

To differentiate between eruptive and confined flares, understanding the physical mechanisms behind flare and coronal mass ejection (CME) initiation is crucial. Solar flares occur when magnetic energy is released and converted into electromagnetic, kinetic, and thermal energy. On the other hand, CMEs happen when plasma trapped within a magnetic flux rope, often visible as a dark filament, is ejected into interplanetary space. This filament, particularly when located above an active region (AR), is significantly influenced by the AR's dynamics. The substantial energy released during a flare can destabilize the flux rope, leading to a CME. However, the filament's release depends on the overlying magnetic field, which can exert a downward force preventing the CME. This instability of the flux rope against its surrounding magnetic field is known as torus instability (TI). In TI analysis, the surrounding magnetic field's condition is represented by the magnetic decay index ($n$), which measures the rate of change of the horizontal magnetic field with height. A flux rope located in a region with a decay index exceeding a certain threshold ($n>1.5$) is considered unstable to torus instability \citep{bateman1978,kliem_2006}.

It is widely believed that electric current flowing in the corona, possibly within a flux rope, plays an important role in the eruptive nature of a flare. The net current flowing in a flux rope may produce a hoop force that expands the current loop, ultimately leading to eruption. This hoop force, counterbalanced by the strapping field, is proportional to the square of the total current (Zakharov \& Shafranov 1986, as cited in \citet{liu_2017}). The equilibrium between these forces determines the stability of the flux rope \citep{kliem_2006,demoulin_2010}. While a single current loop is often considered, \citet{ishiguro_2017} proposed that the current-carrying structure can also be in the form of a double-arc, whose stability is also influenced by its net current. \citet{parker_1996} suggested that the net current, the sum of a direct and return current, of a flux tube is zero. Since an active region (AR) consists of many distinct flux tubes, the entire averaged net current of an AR can also be considered zero. Observations of isolated sunspots showed that the net current from many of these sunspots is near neutrality \citep{venkata_2009}.

\citet{melrose_1991,melrose_1995} argued that the net current of real ARs measured from vector magnetogram data deviates from neutral conditions because non-neutralized currents can be produced with the emergence of magnetic flux during the AR's evolution. \citet{wheatland_2000} further supported this idea by demonstrating that large-scale currents in 21 active regions are predominantly unneutralized. Recent studies have consistently shown a correlation between the degree of current neutralization and flare eruption. ARs producing eruptive flares tend to exhibit non-neutralized currents, while confined flares are often associated with neutralized currents \citep{liu_2017,vemareddy_2019,avallone_2020,liu_2024}. However, it is crucial to note that ARs comprise multiple magnetic systems, and careful consideration must be given to the specific regions relevant to flaring activity when assessing current neutralization.

Previous studies have revealed that the current neutralization measured at the AR scale can differ significantly from that measured within localized region associated with the observed flare \citep{avallone_2020,kazachenko_2022}. To accurately assess the degree of current neutralization, it is essential to apply appropriate localization techniques. This typically involves masking regions based on the magnetic field connectivity of the relevant magnetic structures.

The degree of current neutralization is often correlated with the degree of magnetic shear along the polarity inversion line (PIL) \citep{liu_2017,avallone_2020,kazachenko_2022}. This suggests a link between current neutralization and the non-potentiality of the AR, which may also infer its relation to flaring or eruptive activities. However, although the difference between the current neutralizations of eruptive and confined flare ARs is quite profound in several large flares, a definitive threshold for distinguishing between the two remains elusive. This is due to the overlap between the two groups, making the current neutralization ratio comparable to other proposed parameters in terms of its ability to clearly differentiate between eruptive and confined flares. 

While the torus instability of a flux rope is closely linked to current neutralization in active regions (ARs), the relationship between current neutralization and the decay index of ARs remains unexplored. Understanding this relationship can be important for elucidating the connection between current neutralization and torus instability in ARs. In this study, we investigated the ratios of current neutralization and critical heights of the decay index for flaring ARs that produced and did not produce coronal mass ejections (CMEs). Throughout this paper, the terms non-eruptive and confined flares are used interchangeably. We used vector magnetogram data from the Helioseismic and Magnetic Imager (HMI) instrument \citep{scherrer_2012} on the Solar Dynamics Observatory (SDO) to measure the vertical current densities, derive the coronal potential and nonlinear force-free fields of the ARs, and estimate the critical height of the decay index. We explain details of the data and methods employed in Section 2. In Section 3, we present and discuss the results. Finally, the conclusions are summarized in Section 4.

\section{2.	Data and Methods} \label{sec:datamethod}
\subsection{Data} \label{subsec:data}
We analyzed flaring ARs from the event list of \citet{toriumi_2017} that selected all flares with GOES SXR class over M5.0 within 45° from the disk center in the period from 2010 May to 2016 April.  This event list has also been used by \citet{lin_2020,lin_2021} to study flare eruptivity. We adopted the eruptivity classification from \citet{lin_2020} and further validated it by examining coronal observations from the Large Angle and Spectrometric Coronagraph Experiment (LASCO) instrument \citep{brueckner_1995} on the Solar and Heliospheric Observatory (SOHO), cross-referencing with the Coordinated Data Analysis Workshops (CDAW) CME database $(https://cdaw.gsfc.nasa.gov/CME_list/index.html)$, and comparing with classifications from other studies. After a thorough review, we identified six events that presented inconsistencies. Some flares originating from ARs, such as 11884 and 12241, were classified differently by other studies \citep{yan_2015,wang_2017,joshi_2017}, while others can be considered ambiguous due to absence of LASCO data at that time (AR 11476), identification of suspected related CMEs (M7.2 flare from AR 11944, AR 12403), or inter-ARs flare (X1.2 flare from AR 11944). To maintain clarity and focus on unambiguous events, we excluded these six flares, resulting in a final list of 45 flares (31 eruptive and 14 confined) with confidently assigned categories. List of these flare events is shown in Table \ref{tab:listar}.

We utilized Spaceweather Helioseismic and Magnetic Imager Active Region Patch (SHARP) data, which provides definitive photospheric vector magnetic field components (\texttt{Bp}, \texttt{Bt}, \texttt{Br}) remapped onto a Lambert Cylindrical Equal-Area (CEA) projection \citep{bobra_2014}. We incorporated the error estimates (\texttt{Bp\_err}, \texttt{Bt\_err}, \texttt{Br\_err}) and confidence levels of disambiguation (\texttt{CONF\_DISAMBIG}) provided by SHARP to assess uncertainty and select high-confidence pixels for our analysis, respectively. We selected vector magnetogram data of each flare event within around one hour before the associated flare happened.

To identify relevant magnetic flux structures, we employed nonlinear force-free field (NLFFF) models of the 45 selected ARs from the ISEE NLFFF database hosted by the Institute for Space-Earth Environmental Research, Nagoya University \citep{kusano_2021}. This database contains 3D magnetic field extrapolations, both potential field and NLFFF of ARs analyzed by \citet{kusano_2020}. The NLFFF extrapolations are derived from vector magnetic field observations by the SDO/HMI using the magnetohydrodynamic relaxation method \citep{inoue_2014}.

\begin{deluxetable*}{cccccccc}
\tablenum{1}
\tablecaption{Event List for Analysis \label{tab:listar}}
\tablewidth{1pt}
\tablehead{
\colhead{AR} & \colhead{Date and time} & \colhead{Flare} & \colhead{CME} &
\colhead{Position} & \colhead{$|DC/RC|$} & \colhead{$|DC/RC|$} & \colhead{$|DC/RC|$}  \\
\colhead{(\#) NOAA } & \colhead{(UT)} & \colhead{class} & \colhead{Yes/No} &
\colhead{} & \colhead{$B_{z}>0$} & \colhead{$B_{z}<0$} & \colhead{mean}  
}
\decimalcolnumbers
\startdata
(1) 11158 & 2011-Feb-13T17:28 & M6.6 & Y & S20E05 & $8.66\pm0.14$ & $6.34\pm0.09$ & $7.51\pm1.63$\\
(2) 11158 & 2011-Feb-15T01:44 & X2.2 & Y & S20W10 & $7.96\pm0.08$ & $10.77\pm0.2$ & $9.37\pm1.99$ \\
(3) 11261 & 2011-Aug-3T13:17  & M6.0 & Y & N16W30 & $4.97\pm0.24$ & $4.82\pm0.06$ & $4.90\pm0.11$ \\
(4) 11261 & 2011-Aug-4T03:41  & M9.3 & Y & N16W38 & $1.93\pm0.07$ & $2.38\pm0.04$ & $3.03\pm0.32$ \\
(5) 11283 & 2011-Sep-6T01:35  & M5.3 & Y & N13W07 & $2.59\pm0.04$ & $3.12\pm0.09$ & $2.85\pm0.38$ \\
(6) 11283 & 2011-Sep-6T22:12  & X2.1 & Y & N14W18 & $5.27\pm0.07$ & $5.61\pm0.09$ & $5.44\pm0.24$ \\
(7) 11283 & 2011-Sep-7T22:32  & X1.8 & Y & N14W31 & $3.57\pm0.01$ & $2.05\pm0.04$ & $2.81\pm0.02$ \\
(8) 11402 & 2012-Jan-23T03:38 & M8.7 & Y & N33W21 & $1.92\pm0.04$ & $1.31\pm0.04$ & $1.62\pm0.43$ \\
(9) 11429 & 2012-Mar-7T00:02  & X5.4 & Y & N18E31 & $2.55\pm0.06$ & $2.58\pm0.02$ & $2.56\pm0.03$ \\
(10) 11429 & 2012-Mar-7T01:05  & X1.3 & Y & N15E26 & $2.92\pm0.06$ & $2.62\pm0.03$ & $2.77\pm0.21$ \\
(11) 11429 & 2012-Mar-9T03:22 & M6.3 & Y & N15W03 & $3.16\pm0.16$ & $3.13\pm0.10$ & $3.15\pm0.09$ \\
(12) 11429 & 2012-Mar-10T17:15 & M8.4 & Y & N17W24 & $2.26\pm0.05$ & $2.70\pm0.07$ & $2.48\pm0.31$ \\
(13) 11515 & 2012-Jul-2T10:43  & M5.6 & Y & S17E06 & $1.08\pm0.05$ & $2.55\pm0.04$ & $1.81\pm1.04$ \\
(14) 11520 & 2012-Jul-12T15:37  & X1.4 & Y & S13W03 & $2.11\pm0.07$ & $1.66\pm0.01$ & $1.89\pm0.32$ \\
(15) 11719 & 2013-Apr-11T06:55  & M6.5 & Y & N07E13 & $1.17\pm0.04$ & $1.79\pm0.08$ & $1.48\pm0.44$ \\
(16) 11877 & 2013-Oct-24T00:21  & M9.3 & Y & S09E10 & $2.51\pm0.06$ & $4.14\pm0.11$ & $3.32\pm1.15$ \\
(17) 11890 & 2013-Nov-5T22:07  & X3.3 & Y & S12E44 & $4.57\pm0.43$ & $2.84\pm0.11$ & $3.71\pm1.23$ \\
(18) 11890 & 2013-Nov-8T04:20 & X1.1 & Y & S13E13 & $6.62\pm0.32$ & $4.80\pm0.03$ & $5.71\pm1.29$ \\
(19) 11890 & 2013-Nov-10T05:08  & X1.1 & Y & S13W13 & $10.70\pm0.25$ & $7.93\pm0.39$ & $9.31\pm1.97$ \\
(20) 11936 & 2013-Dec-31T21:45  & M6.4 & Y & S15W36 & $1.29\pm0.02$ & $1.24\pm0.03$ & $1.27\pm0.04$ \\
(21) 12017 & 2014-Mar-29T17:35 & X1.0 & Y & N10W32 & $2.54\pm0.02$ & $2.11\pm0.04$ & $2.32\pm0.31$ \\
(22) 12036 & 2014-Apr-18T12:31 & M7.3 & Y & S20W34 & $2.79\pm0.19$ & $2.38\pm0.12$ & $2.59\pm0.29$ \\
(23) 12158 & 2014-Sep-10T17:21  & X1.6 & Y & N11E05 & $1.92\pm0.03$ & $1.83\pm0.05$ & $1.88\pm0.06$ \\
(24) 12173 & 2014-Sep-28T02:39  & M5.1 & Y & S13W23 & $2.21\pm0.07$ & $1.01\pm0.04$ & $1.61\pm0.85$ \\
(25) 12205 & 2014-Nov-7T16:53  & X1.6 & Y & N17E40 & $3.21\pm0.08$ & $3.14\pm0.09$ & $3.18\pm0.06$ \\
(26) 12242 & 2014-Dec-17T04:25  & M8.7 & Y & S18E08 & $1.80\pm0.05$ & $1.45\pm0.03$ & $1.63\pm0.25$ \\
(27) 12242 & 2014-Dec-20T00:11  & X1.8 & Y & S19W29 & $1.52\pm0.02$ & $2.16\pm0.13$ & $1.84\pm0.45$ \\
(28) 12297 & 2015-Mar-10T03:19 & M5.1 & Y & S15E39 & $1.61\pm0.04$ & $1.92\pm0.05$ & $1.77\pm0.22$ \\
(29) 12297 & 2015-Mar-11T16:11  & X2.1 & Y & S17E22 & $1.98\pm0.05$ & $1.86\pm0.004$ & $1.92\pm0.08$ \\
(30) 12371 & 2015-Jun-22T17:39  & M6.5 & Y & N13W06 & $3.07\pm0.09$ & $2.17\pm0.07$ & $2.62\pm0.64$ \\
(31) 12371 & 2015-Jun-25T08:02 & M7.9 & Y & N12W40 & $2.92\pm0.04$ & $1.37\pm0.04$ & $2.15\pm1.09$ \\
\hline
(32) 11166 & 2011-Mar-9T23:13 & X1.5 & N & N08W11 & $1.29\pm0.03$ & $1.40\pm0.04$ & $1.34\pm0.08$ \\
(33) 11261 & 2011-Jul-30T02:04  & M9.3 & N & N14E35 & $1.34\pm0.01$ & $1.48\pm0.06$ & $1.41\pm0.10$ \\
(34) 11515 & 2012-Jul-4T09:47  & M5.3 & N & S17W18 & $1.02\pm0.02$ & $1.08\pm0.03$ & $1.05\pm0.04$  \\
(35) 11515 & 2012-Jul-5T11:39  & M6.1 & N & S18W32 & $1.07\pm0.01$ & $1.09\pm0.02$ & $1.08\pm0.01$ \\
(36) 11884 & 2013-Nov-3T05:16  & M5.0 & N & S12W17 & $1.83\pm0.05$ & $3.06\pm0.11$ & $2.44\pm0.87$ \\
(37) 11967 & 2014-Feb-4T03:57  & M5.2 & N & S14W07 & $2.32\pm0.10$ & $1.43\pm0.34$ & $1.88\pm0.63$ \\
(38) 12192 & 2014-Oct-22T01:16 & M8.7 & N & S13E21 & $1.21\pm0.03$ & $1.13\pm0.02$ & $1.17\pm0.06$ \\
(39) 12192 & 2014-Oct-22T14:02  & X1.6 & N & S14E13 & $1.16\pm0.03$ & $1.14\pm0.03$ & $1.15\pm0.02$\\
(40) 12192 & 2014-Oct-24T21:07  & X3.1 & N & S22W21 & $1.12\pm0.003$ & $1.30\pm0.03$ & $1.21\pm0.13$ \\
(41) 12192 & 2014-Oct-25T16:55 & X1.0 & N & S10W22 & $1.19\pm0.02$ & $1.32\pm0.02$ & $1.26\pm0.09$ \\
(42) 12192 & 2014-Oct-26T10:04 & X2.0 & N & S14W37 & $1.24\pm0.03$ & $1.20\pm0.02$ & $1.22\pm0.03$ \\
(43) 12192 & 2014-Oct-27T00:06  & M7.1 & N & S12W42 & $1.16\pm0.01$ & $1.07\pm0.01$ & $1.11\pm0.06$\\
(44) 12222 & 2014-Dec-4T18:05  & M6.1 & N & S20W31 & $1.17\pm0.02$ & $1.51\pm0.07$ & $1.34\pm0.25$ \\
(45) 12422 & 2015-Sep-28T14:53  & M7.6 & N & S20W28 & $1.07\pm0.15$ & $1.70\pm0.13$ & $1.39\pm0.44$\\
\enddata
\tablecomments{Horizontal line separates between eruptive (above) and confined flares (below).}
\end{deluxetable*}

\subsection{Selection of Relevant Magnetic Fluxes } \label{subsec:selection}
\begin{figure}[ht!]
\plotone{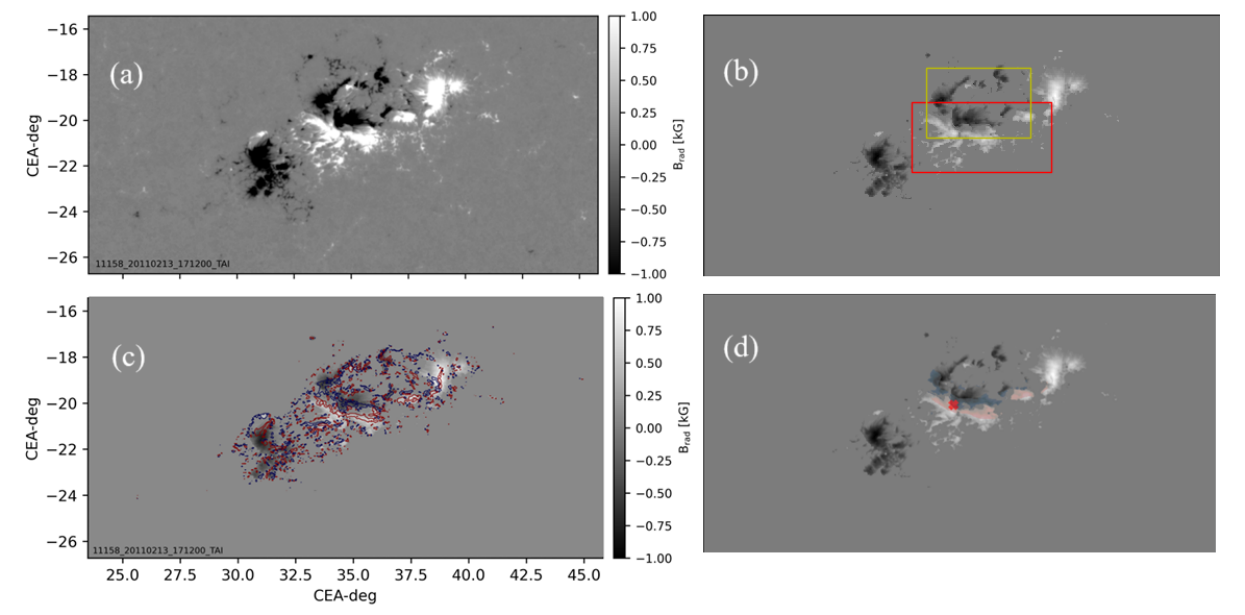}
\caption{(a) Vertical component of vector magnetic field of AR 11158. (b) Red (yellow) rectangle outlines boundary of ROI where field lines are traced from positive (negative) polarity region. (c) Vertical component of current density derived from the vector magnetogram data. Red and blue contours mark $J_z$ of 20 and -20 mA/$m^{2}$. (d) Contours of magnetic flux used in the analysis, shaded by red (blue) for positive (negative) polarities. Red crossmark indicates location of the maximum free-energy density point.   \label{fig:fig1}}
\end{figure}

To identify the magnetic flux structures relevant to flaring activity in each AR, we first determined the flare location using Atmospheric Imaging Assembly (AIA) 1600\textup{~\AA} observations from SDO. We then examined the NLFFF model of the corresponding AR by tracing magnetic field lines from pixels of one polarity to the opposite polarity within a region of interest (ROI), which was manually predefined based on AIA observations. For clarity, we show famous AR 11158 on 2011 February 13 at 17.12 UT as an example for selecting the relevant flux in our method (see Figure \ref{fig:fig1}). Red rectangle in Figure \ref{fig:fig1}b marks the region where we started to trace field lines from positive polarity regions. By identifying the endpoints of each field line within the target ROI (yellow rectangle), we were able to exclude field lines connecting to irrelevant regions outside the ROI. Red and blue shaded regions in Figure \ref{fig:fig1}d show the selected masks for the relevant magnetic fluxes included in the analysis. This process allowed us to isolate the magnetic flux directly involved in the flaring activity. 

Our method for selecting relevant magnetic flux is more straightforward than those employing the squashing factor Q-map \citep{liu_2017,liu_2024}. Q-map essentially highlighting regions of rapid change in field-line connectivity \citep{titov_2002,titov_2007}. However, Q-maps can be complex to interpret, requiring additional steps to filter out irrelevant magnetic flux. Therefore, our method offers a simpler approach that directly identifies the relevant magnetic flux.

\subsection{Calculation of Current Neutralization} \label{subsec:calculation}
To quantify the degree of current neutralization of an AR, we first calculated vertical component of the current density, $J_z$, derived from the horizontal components of vector magnetogram data provided by SHARP, using Ampere’s law:
\begin{equation}
J_z = \frac{1}{\mu_0} (\frac{\partial B_y}{\partial x} - \frac{\partial B_x}{\partial y}).
\end{equation}
Figure \ref{fig:fig1}c shows distribution of vertical component of current density of AR 11158. 

Subsequently, we isolate and separate current from the positive and negative polarity regions. To accurately calculate the degree of current neutralization, we selected only high-confidence pixels identified by SHARP with a \texttt{CONF\_DISAMBIG} value of 90. This approach excludes pixels with significant uncertainty, allowing us to incorporate all magnetic field values from the magnetogram that exhibit high confidence levels.

Following the pixels selection, we calculated the direct current (DC) and return current (RC) for each polarity.  This involved integrating $J_z$ with opposite directions ($J_z^+$ or $J_z^-$) over selected regions in positive and negative polarities. Association of the correct sign of $J_z$ with DC and RC was determined by the dominant helicity \citep{liu_2017,kazachenko_2022,liu_2024}. It was conducted by integrating $J_zB_z$ over the ROI as the proxy for current helicity. Consequently, for a region with dominant positive helicity, the $DC$ and $RC$ in the positive and negative polarities are defined as:

\begin{equation}
	\begin{aligned}	
			DC^+ = \int {J_z^+ ds^+} , && RC^+ = \int {J_z^- ds^+}  \\
			DC^- = \int {J_z^- ds^-} , && RC^- = \int {J_z^+ ds^-} &&.
	\end{aligned} 
\end{equation}
For a region with dominant negative helicity,  
\begin{equation}
	\begin{aligned}	
			DC^+ = \int {J_z^- ds^+} , && RC^+ = \int {J_z^+ ds^+}  \\
			DC^- = \int {J_z^+ ds^-} , && RC^- = \int {J_z^- ds^-} &&.
	\end{aligned} 
\end{equation}

The degree of current neutralization of each polarity was calculated as $|DC/RC|^+$ and $|DC/RC|^-$, respectively. The total degree of current neutralization $|DC/RC|$ was defined as the average of these values for both polarities \citep{dalmasse_2015, kazachenko_2022, liu_2017, liu_2024}. 
To estimate the uncertainty in $|DC/RC|^+$and $|DC/RC|^-$, we used error values of the components of vector magnetic fields provided by SHARP data, rather than  using a constant error value as in other studies \citep{liu_2024,vemareddy_2019}. \citet{vemareddy_2019} showed that a constant horizontal field error of 40 G leads to a maximum uncertainty in $|DC/RC|$ of 0.14, which is relatively small compared to the evolution of $|DC/RC|$ during pre-flare activities. For the total degree of current neutralization $|DC/RC|$ error, we followed the approach of \citet{liu_2024} by comparing the absolute errors of $|DC/RC|^+$ and $|DC/RC|^-$ with the standard deviation of $|DC/RC|$. The larger of these two values was adopted as the final error estimate.

\subsection{Calculation of critical height} \label{subsec:critic}

Critical height of a flux rope to experience torus instability is commonly defined as the height where magnetic decay index in a volumetric coronal field reaches the threshold of 1.5. Usually, decay index is derived from the 3D potential field data extrapolated from the radial component of vector magnetogram data. However, precisely identifying and tracking flux ropes in 3D NLFFF models remains challenging. This difficulty arises from the often ambiguous nature of flux rope structures in such models. Furthermore, accurately estimating the height of a filament, a potential manifestation of a flux rope, is hindered by the requirement for multiple observations from different point of views. As a result, determining the exact critical height associated with a flare-related flux rope remains a complex task.

In this study, we use a simple approach to define critical height as the height above a specific photospheric point exhibiting the maximum photospheric magnetic free-energy density within the ROI of each AR. To locate this point, we estimated the photospheric magnetic free-energy density at each pixel by calculating the proxy of magnetic free-energy density as 
\begin{equation}
E_f = \frac{(\boldsymbol{B_{obs}}-\boldsymbol{B_{pot}})^2}{8\pi}.
\end{equation}

Here, $\boldsymbol{B_{obs}}$ is the tangential component of the observed vector magnetic fields ($B_x$ and $B_y$) provided by SHARP, while $\boldsymbol{B_{pot}}$ is the tangential component of the potential field, derived from the vertical component of magnetogram $B_z$ using the Fourier method. Subsequently, we search the coordinate of the maximum value of this proxy of photospheric magnetic free-energy density in the magnetogram data. We found that these maximum free-energy density points were consistently located near the polarity inversion line (PIL), specifically between the flare ribbon locations. These points are assumed to be located beneath the cores of the flux ropes.   

To calculate decay index, we used 3D potential field data from ISEE NLFFF database. The decay index at each grid point in the volumetric space was calculated as:

\begin{equation}
n = -\frac{z}{B_{p}}\frac{\partial B_{p}}{\partial z},
\end{equation}
where $z$ is the height from the photosphere and $B_{p}$ denotes horizontal component of the potential field \citep{Inoue2018}.  

Given potential discrepancies in the dimensional size between the bottom boundary of the 3D potential field and the original magnetogram, we projected the coordinates of the maximum free-energy density point onto the extrapolated data using a proportional scaling factor. A vertical profile of the decay index was then generated above this maximum free-energy density point. The critical height along this profile was defined as the minimum height at which the decay index reached the threshold value of 1.5. To enhance statistical robustness, we also derived multiple decay index profiles from the surrounding region of the maximum free-energy density point. The mean critical height from these profiles was considered the representative critical height for the AR. The uncertainty in the critical height was estimated from the standard deviation of the critical heights obtained from the maximum free-energy density point and its surrounding points.

\section{3.	Results and Discussions} \label{sec:res}
\subsection{Electric current neutralizations distributions} \label{subsec:currentdis}

\begin{figure}[ht!]
\plotone{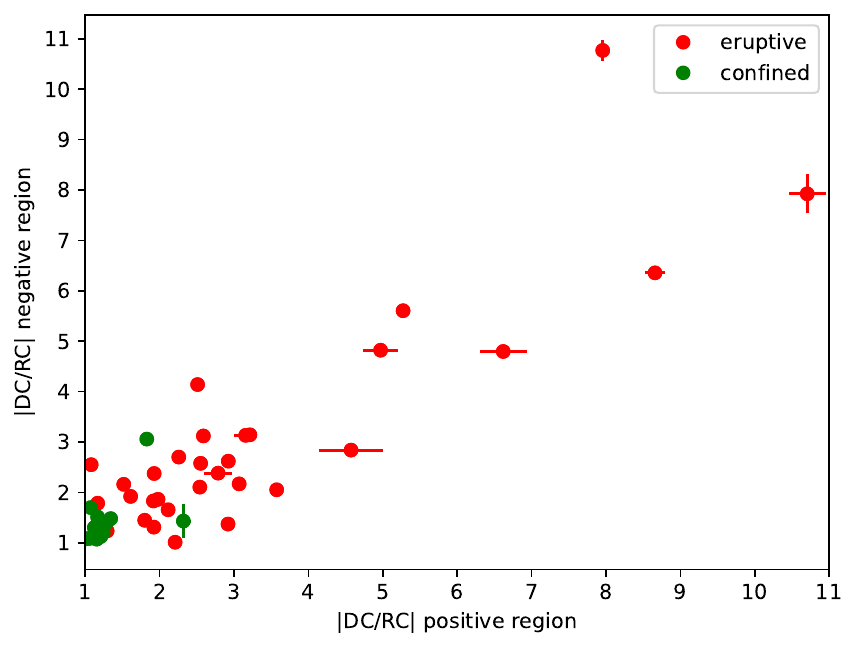}
\caption{Degrees of current neutralization from 45 flare events calculated for different magnetic polarities.  \label{fig:fig2}}
\end{figure}

Figure \ref{fig:fig2} shows plot of $|DC/RC|$ calculated for both positive and negative polarity regions listed in columns (6) and (7) of Table \ref{tab:listar}. While generally consistent, slight discrepancies can be observed between the two polarities. These differences may arise from limitations in the NLFFF model, which may not accurately represent the true coronal magnetic field lines, potentially leading to the selection of the non-ideal fluxes related to the flux ropes. The total degree of current neutralizations for all events are provided in column (8) of Table \ref{tab:listar}. We can clearly see that there is a different pattern between eruptive and non-eruptive flares. Non-eruptive flares tend to have $|DC/RC|$ close to unity in both polarities, suggesting a high degree of current neutralization in these flaring ARs. On the other hand, most eruptive flares exhibit non-neutralized current conditions. These results are consistent with previous studies that have reported similar characteristics between these two types of flares. 

\begin{figure}[ht!]
\plotone{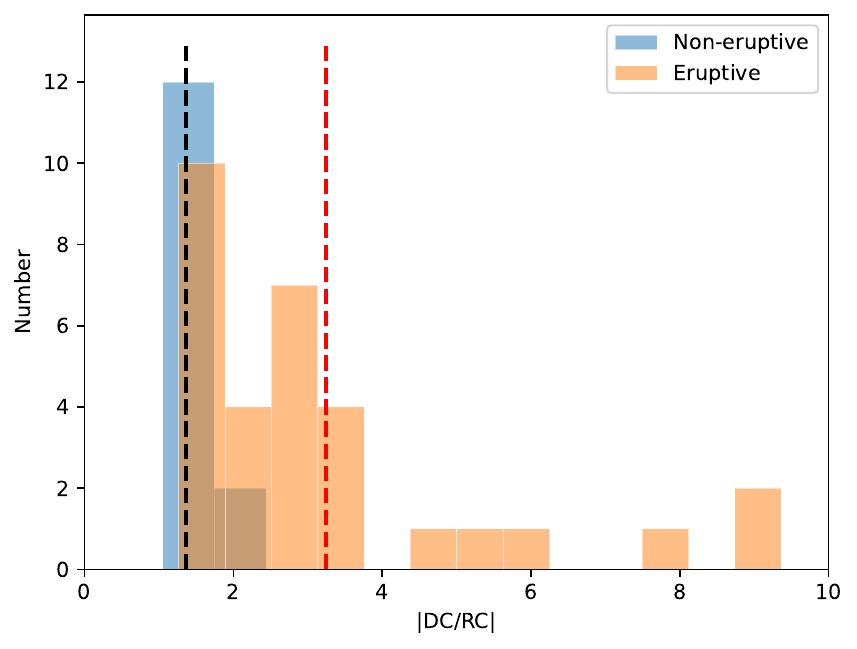}
\caption{Distribution of $|DC/RC|$ for eruptive and non-eruptive flares. Dash vertical lines mark the mean value of each population.  \label{fig:fig3}}
\end{figure}

We plotted the distribution of $|DC/RC|$ in Figure \ref{fig:fig3}. The results indicate that most non-eruptive flares exhibit near-neutral current conditions, with values ranging from 1.05 to 2.44 and an average of 1.36. This value is notably smaller than the average value of 2.2 reported for confined flares by \citet{liu_2024}. On the other hand, eruptive flares generally display non-neutralized current conditions, with values ranging from 1.27 to 9.37 and an average of 3.24. This average value is comparable to the value of 3.6 reported for X-class eruptive flares by \citet{liu_2024}. These values are significantly larger than $|DC/RC|$ values derived from larger integration area, e.g., in \citet{avallone_2020}, that calculated $|DC/RC|$ without selectively chose integration area using connectivity information, such as from NLFFF model. Our results consistent with previous studies that have demonstrated that $|DC/RC|$ calculated within smaller ROI is typically higher than values obtained from larger regions or the entire AR \citep{kazachenko_2022}.
	 
While our findings generally indicate that eruptive flares are associated with non-neutralized current conditions, we also identified quite significant events that are close to neutrality. This observation aligns with other studies, suggesting that relying solely on the degree of current neutralization may not be sufficient for distinguishing between eruptive and non-eruptive flares. However, the degree of current neutralization can provide insights into the non-potentiality of the AR, which can also be approximated by other parameters such as the shear angle near the polarity inversion line (PIL), as demonstrated by previous researches \citep{avallone_2020,kazachenko_2022,liu_2017,liu_2024}.

\subsection{Critical heights} \label{subsec:critdis}
\begin{figure}[ht!]
\includegraphics[scale=0.8]{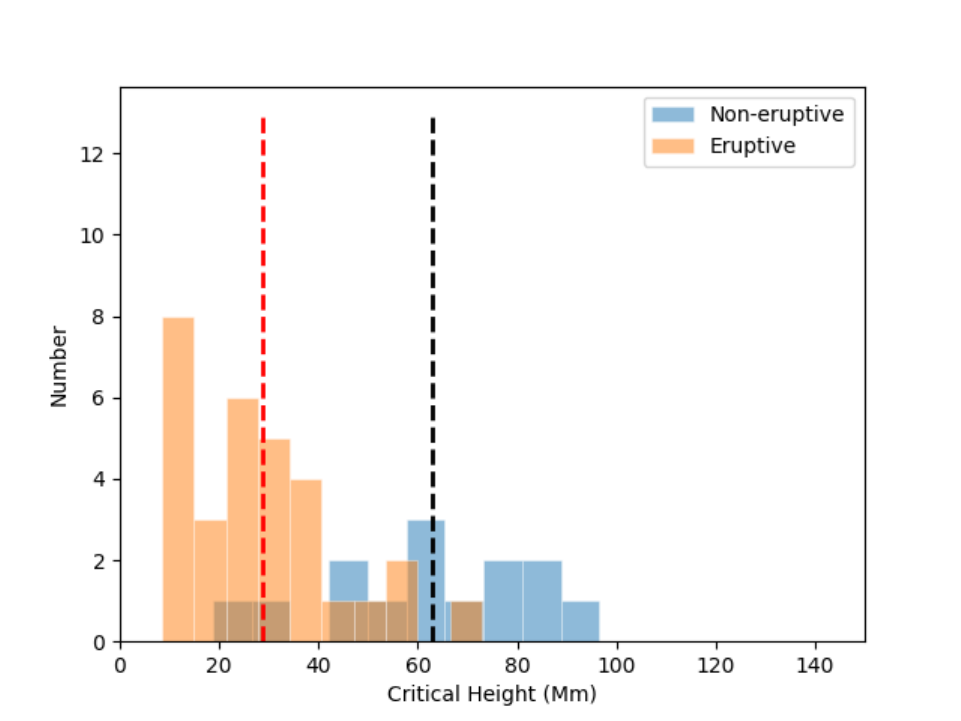}
\centering
\caption{Distribution of critical heights for eruptive and non-eruptive flares. Dash vertical lines mark the mean value of each population. \label{fig:fig4}}
\end{figure}

Figure \ref{fig:fig4} depicts the distribution of critical heights for different flare categories. Eruptive events generally exhibit lower critical heights compared to non-eruptive events. The critical heights for eruptive events range from 8.43 to 72.90 Mm, with an average of 28.74 Mm. In contrast, non-eruptive events have critical heights ranging from 18.73 to 96.60 Mm, with an average of 62.85 Mm. The complete list of critical heights for all the flare events is shown in column (2) of Table \ref{listcrit}. These results are consistent with the expectation that flux ropes associated with eruptive events are more likely to attain critical heights, leading to CME production. Conversely, confined flares may struggle to reach these critical heights, preventing CME initiation.

Similar to the distribution of the degree of current neutralization, there is significant overlap between the two categories. This suggests that relying solely on critical height may not be sufficient for reliably distinguishing between eruptive and non-eruptive flares.

\begin{deluxetable*}{cccc}
\tablenum{2}
\tablecaption{Critical Heights and S-parameters of all the flare events \label{listcrit}}
\tablewidth{1pt}
\tablehead{
\colhead{\#} & \colhead{Critical height} & \colhead{nch} & \colhead{S}  \\
\colhead{} & \colhead{(Mm)} & \colhead{} & \colhead{} 
}
\decimalcolnumbers
\startdata
1 & $9.62\pm0.59$ & $0.24\pm0.015$ & $30.72\pm6.92$\\
2 & $39.94\pm0.63$ & $1.01\pm0.02$& $9.23\pm1.967$\\
3 & $23.42\pm0.67$ & $0.60\pm0.017$& $8.22\pm0.31$\\
4 & $17.82\pm10.66$ & $0.46\pm0.27$& $6.68\pm4.0$\\
5 & $11.87\pm0.42$& $0.30\pm0.01$& $9.47\pm1.29$\\
6 & $10.68\pm0.42$& $0.27\pm0.01$& $20.02\pm1.18$\\
7 & $12.06\pm0.61$& $0.31\pm0.016$& $9.17\pm3.54$\\
8 & $38.74\pm1.59$& $0.98\pm0.04$& $1.64\pm0.45$\\
9 & $33.28\pm8.17$& $0.85\pm0.21$& $3.03\pm0.74$\\
10 & $31.51\pm11.16$ & $0.80\pm0.28$& $3.46\pm1.26$\\
11 & $57.66\pm0.0.67$ & $1.47\pm0.017$& $2.15\pm0.07$\\
12 & $32.69\pm1.36$& $0.83\pm0.035$& $2.99\pm0.40$\\
13 & $22.42\pm0.56$& $0.57\pm0.014$& $3.19\pm1.83$\\
14 & $26.80\pm10.83$ & $0.68\pm0.02$& $2.77\pm0.48$\\
15 & $8.43\pm5.80$& $0.21\pm0.15$& $6.89\pm5.16$\\
16 & $30.21\pm1.22$ & $0.77\pm0.031$& $4.32\pm1.51$\\
17 & $10.70\pm1.40$ & $0.27\pm0.03$& $13.62\pm4.85$\\
18 & $37.03\pm6.96$ & $0.94\pm0.77$& $6.07\pm5.15$\\
19 & $58.53\pm16.23$ & $1.49\pm0.41$& $6.26\pm2.18$ \\
20 & $25.53\pm1.42$ & $0.65\pm0.036$ & $1.95\pm0.12$\\
21 & $12.67\pm0.0.85$ & $0.32\pm0.02$ & $7.20\pm1.07$\\
22 & $37.98\pm1.33$ & $0.97\pm0.034$ & $2.68\pm0.313$ \\
23 & $29.93\pm0.60$ & $0.76\pm0.015$& $2.47\pm0.09$\\
24 & $27.73\pm0.85$ & $0.70\pm0.02$& $2.28\pm1.20$\\
25 & $72.90\pm0.90$ & $1.85\pm0.023$& $1.72\pm0.04$\\
26 & $20.11\pm1.11$ & $0.51\pm0.028$& $3.18\pm0.52$\\
27 & $44.49\pm4.02$ & $1.13\pm0.10$& $1.62\pm0.432$\\
28 & $17.17\pm1.61$ & $0.44\pm0.041$& $4.04\pm0.63$\\
29 & $12.82\pm3.83$ & $0.33\pm0.097$& $5.89\pm1.78$\\
30 & $25.86\pm1.80$ & $0.66\pm0.046$& $3.99\pm1.01 $\\
31 & $50.39\pm1.21$ & $1.28\pm0.03$& $1.68\pm0.86$\\
\hline
32 & $47.77\pm0.69$ & $1.21\pm0.02$& $1.11\pm0.90$\\
33 & $32.05\pm0.52$ & $0.81\pm0.013$ & $1.73\pm0.12$\\
34 & $83.22\pm17.80$ & $2.11\pm0.45$& $0.50\pm0.11$\\
35 & $75.73\pm18.22$ & $1.92\pm0.46$& $0.56\pm0.13 $\\
36 & $61.37\pm0.93$ & $1.56\pm0.02$& $1.56\pm0.56$\\
37 & $18.73\pm1.36$ & $0.48\pm0.03$& $3.94\pm1.35$\\
38 & $56.18\pm13.62$ & $1.43\pm0.35$& $0.82\pm0.20$\\
39 & $63.55\pm1.88$ & $1.62\pm0.048$& $0.71\pm0.02$\\
40 & $63.85\pm22.52$ & $1.62\pm0.57$& $0.75\pm0.27$\\
41 & $80.47\pm1.43$ & $2.04\pm0.036$& $0.61\pm0.05$\\
42 & $83.67\pm0.93$ & $2.13\pm0.02$& $0.57\pm0.02$\\
43 & $96.60\pm2.24$ & $2.45\pm0.06$& $0.45\pm0.03 $\\
44 & $48.46\pm1.02$ & $1.23\pm0.03$& $1.09\pm0.20$\\
45 & $68.24\pm14.09$ & $1.73\pm0.36$& $0.80\pm0.30$\\
\enddata
\tablecomments{Horizontal line separates between eruptive (above) and confined flares (below).}
\end{deluxetable*}

\subsection{Current neutrality vs Critical height} \label{subsec:currvscrit}

\begin{figure}
\includegraphics[scale=0.8]{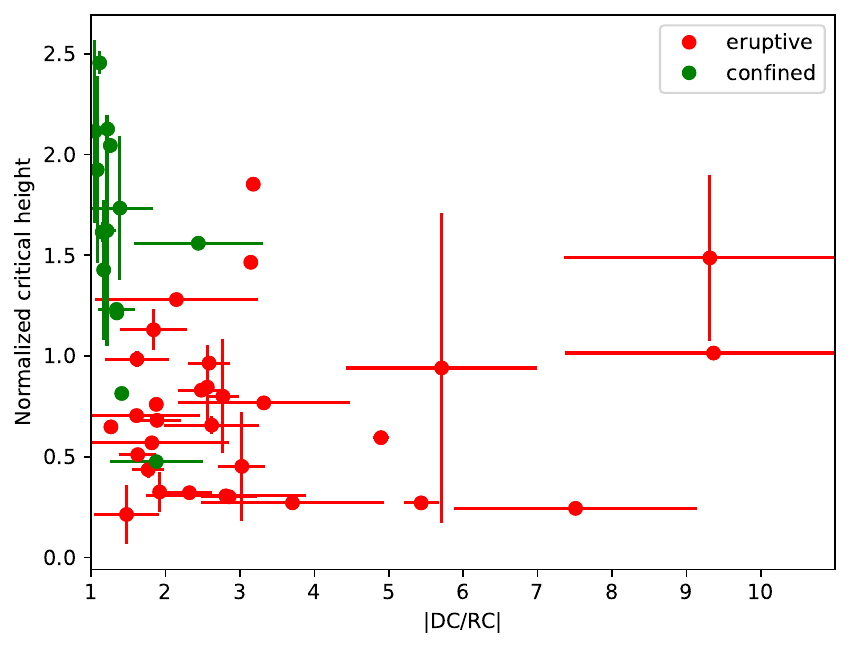}
\centering
\caption{Plot of critical heights of decay index and degrees of current neutralization for eruptive and non-eruptive flares. Critical heights are normalized by mean value of all critical heights in the dataset.   \label{fig:fig5}}
\end{figure}

\begin{figure}
\includegraphics[scale=0.8]{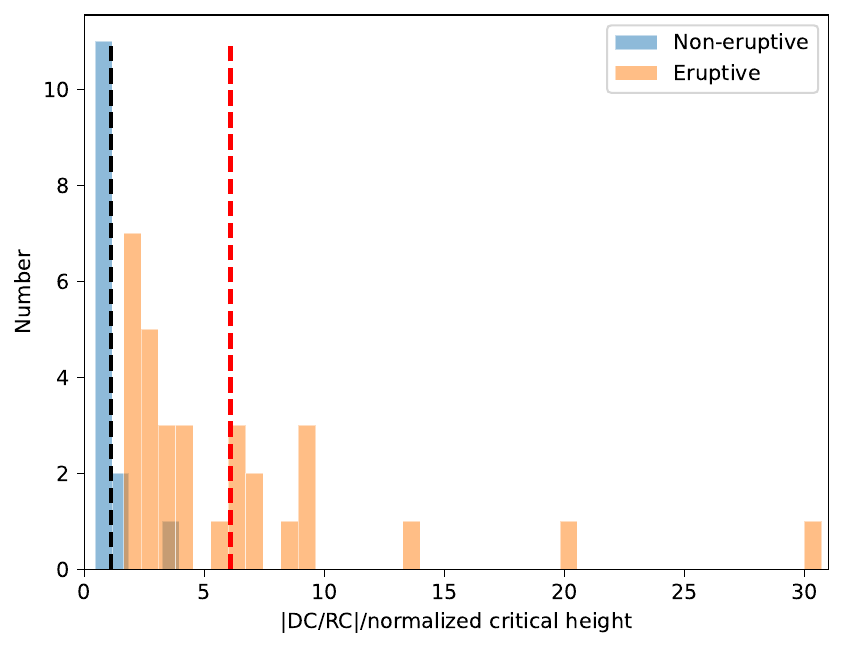}
\centering
\caption{Distribution of S-parameter for eruptive and non-eruptive flares. Dash vertical lines mark the mean value of each population. \label{fig:fig6}}
\end{figure}

It has been shown that $|DC/RC|$ and critical height can be generally used to differentiate between eruptive and non-eruptive events. Figure \ref{fig:fig5} shows the plot of these two parameters for the 45 events in our dataset. This figure clearly demonstrate that eruptive events tend to originate from ARs that are current non-neutralized or have low critical heights. Conversely, confined flares are typically characterized by neutralized current conditions and high critical heights. These findings highlight the value of combining information on both the degree of current neutralization and critical height for discriminating between eruptive and non-eruptive events.

As illustrated in the figure, eruptive flares can occur even at relatively high critical heights, provided the AR exhibits non-neutralized current conditions. On the other hand, eruptions can also occur in ARs with near-neutral current conditions but low critical heights. Our results suggest that the degree of current neutralization, which is calculated from a selected region associated with a flaring activity, can be interpreted as the representation of net current within the flux rope. Thus, it may primarily reflect the hoop force acting on the flux rope, without fully capturing the influence of the overlying field that counteracts this force. Incorporating information on the critical height, which serves as a proxy for the overlying field, can provide a more comprehensive assessment of flux rope stability or instability. Although critical height of decay index is not the direct measure of torus instability condition, it provides valuable insights into the stability of a pre-existing flux rope. By examining the critical height above the core region, we can assess the likelihood of the flux rope becoming unstable.

Furthermore, we introduce a new non-dimensional parameter, 
\begin{equation}
S = \frac{|DC/RC|}{nch},
\end{equation}

where the normalized critical height ($nch$) is obtained by dividing the critical height by the mean critical height value of 39.35 Mm. List of $nch$ and S values for all the events is shown in columns (3) and (4) of Table \ref{listcrit}. The distribution of S for all the events, shown with different colors for categories, is shown in Figure \ref{fig:fig6}. The mean S value for non-eruptive events is 1.09, while for eruptive events, it is 6.08. Unlike the distributions of the degree of current neutralization and critical height, the distribution of S exhibits minimal overlap between eruptive and non-eruptive events. This suggests that the S-parameter has potential as a robust discriminator between these two categories. However, a few ARs exhibit deviations from this general trend. We found that there is no eruptive event with S value lower than 1.63. On the other hand, two confined flare events have S values larger than 1.63. They are event \#33 (AR 11261) and \#37 (AR 11967) with S values of 1.73  and 3.94, respectively. These ARs will be discussed further in the following subsections.

\subsection{Special events} \label{subsec:special}

\subsubsection{AR 11261} \label{subsec:ar11261}
\begin{figure}
\gridline{\fig{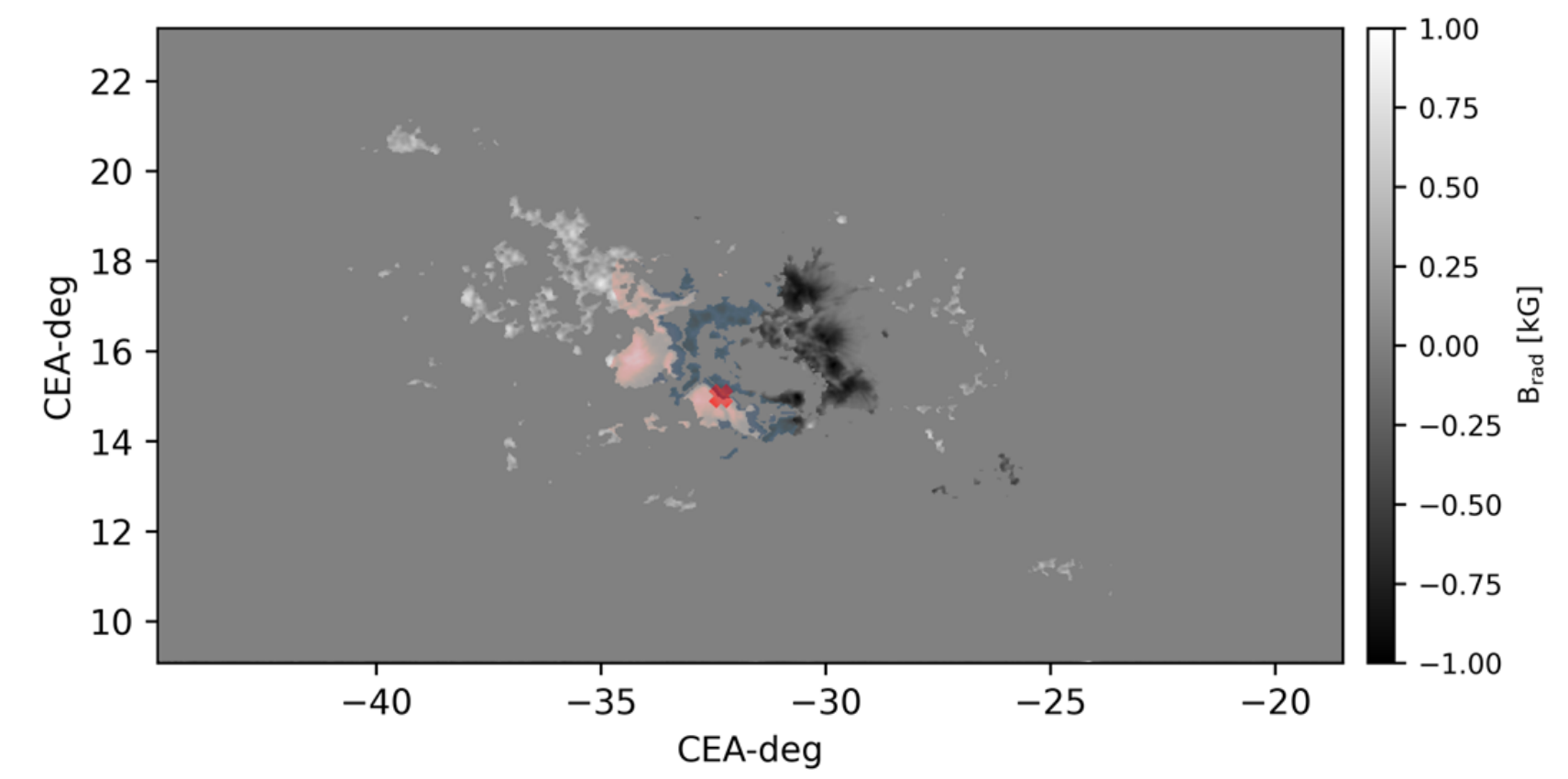}{0.9\textwidth}{(a)}
          }
\gridline{\fig{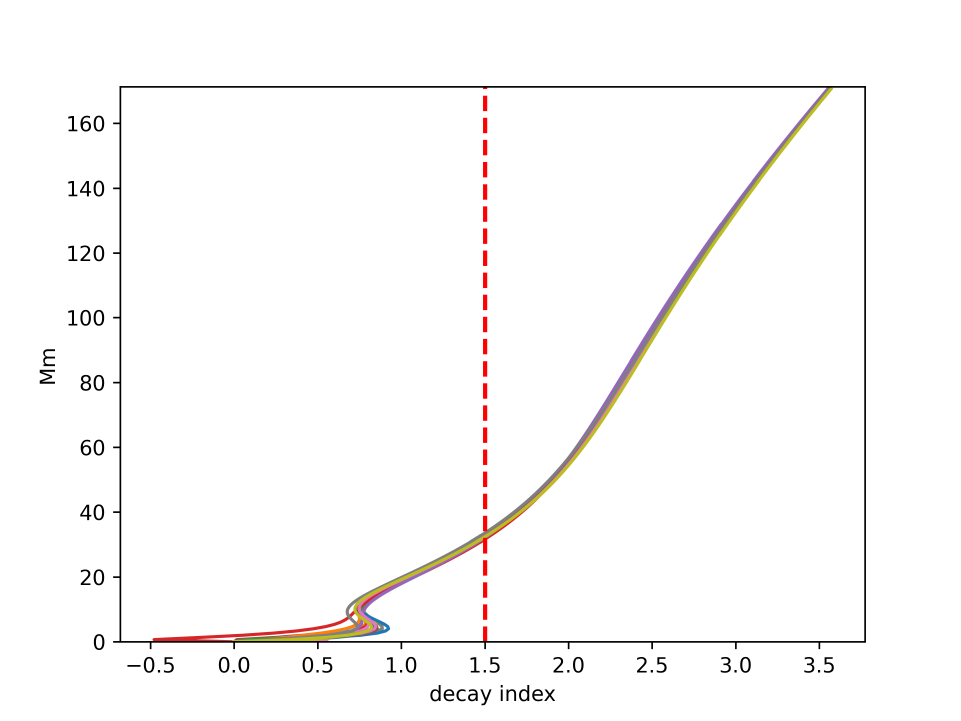}{0.5\textwidth}{(b)}
          }
\caption{Masked regions of AR 11261 where magnetic fluxes used to analyze degree of current neutralization (a). Red (blue) shaded regions mark the relevant fluxes at positive (negative) polarities. Red cross mark shows the location of maximum free-energy density point. Decay index profile along vertical line above the maximum free-energy density point (b). Dash vertical line marks the threshold of torus instability.
\label{fig:fig7}}
\end{figure}
Figure \ref{fig:fig7} shows relevant magnetic flux (a) and decay index profile (b) of AR 11261. For the case of AR 11261, with $|DC/RC|$ of $1.41 \pm 0.097$ and a critical height of $32.05 \pm 0.52$ Mm, the conditions are closer to neutrality compared to most eruptive events. However, the critical height is still below the average of the mean of all critical heights, which is around 40 Mm. The decay index profile for AR 11261 exhibits no anomalies (Figure \ref{fig:fig7}b). Our results suggest that, for this flare event, the values of the degree of current neutralization and critical height fall within a transitional range between eruptive and non-eruptive thresholds. This may indicate that the AR was in a metastable state, in which the AR may easily transform from non-eruptive to eruptive condition. Indeed, this AR produced several CMEs over several days, some of which are included in Table 1 (events \#3 and \#4). These findings again highlight the potential of combining the degree of current neutralization and critical height to discriminate between eruptive and non-eruptive behavior, although definitive discrimination remains challenging, especially for ARs in transitional states.

\subsubsection{AR 11967} \label{subsec:ar11967}

\begin{figure}
\gridline{\fig{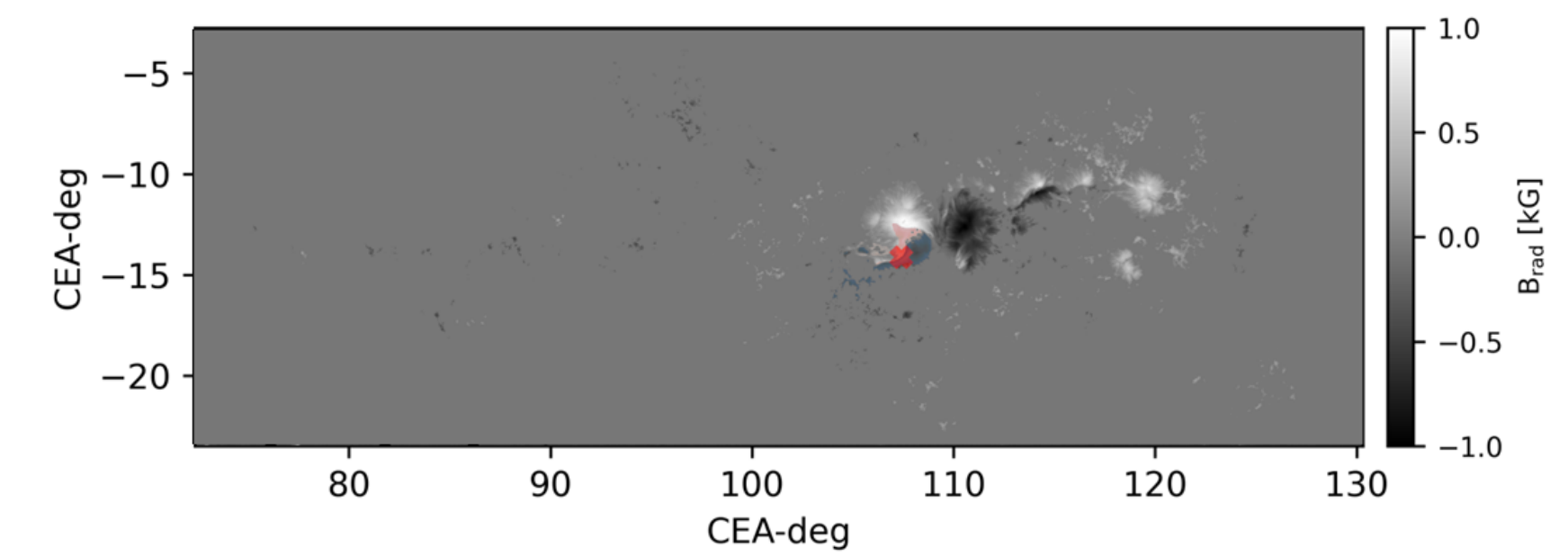}{0.9\textwidth}{(a)}
          }
\gridline{\fig{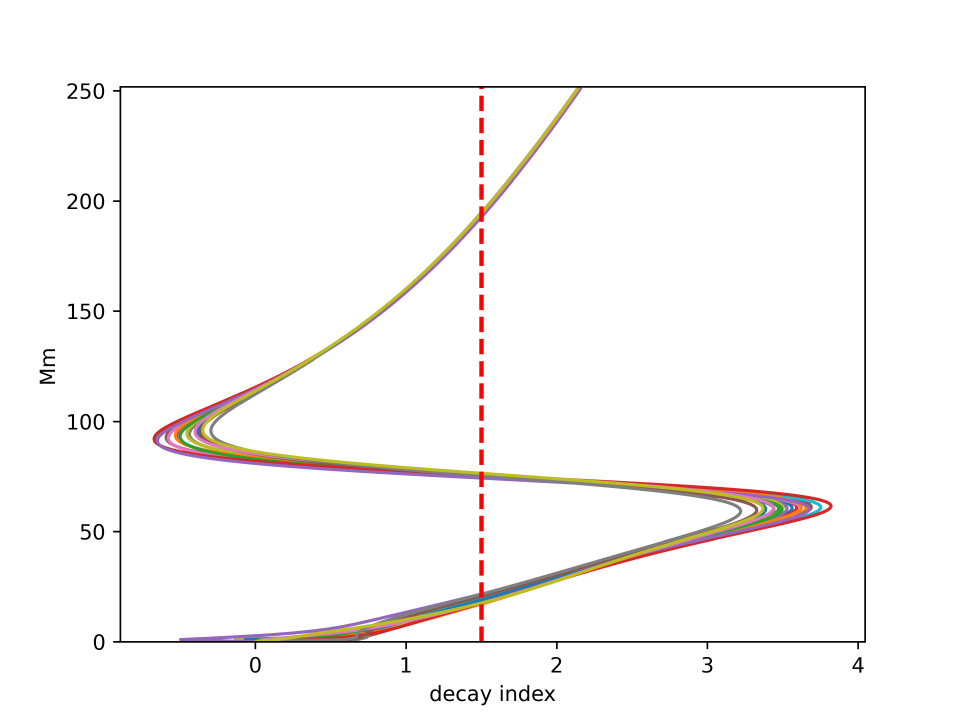}{0.5\textwidth}{(b)}
          }
\caption{Masked regions of AR 11967 where magnetic fluxes used to analyze degree of current neutralization (a). Red (blue) shaded regions mark the relevant fluxes at positive (negative) polarities. Red cross mark shows the location of maximum free-energy density point. Decay index profile along vertical line above the maximum free-energy density point (b). Dash vertical line marks the threshold of torus instability. 
\label{fig:fig8}}
\end{figure}

AR 11967 has $|DC/RC|$ of $1.88 \pm 0.63$, and critical height of $18.73 \pm 1.36$ Mm. Figure \ref{fig:fig8} shows relevant magnetic flux (a) and decay index profile (b) of AR 11967. From Figure \ref{fig:fig8}b, we can see that AR 11967 exhibits a special profile of decay index. The decay index rapidly increases to the critical threshold at a low altitude but then decreases significantly around 70 Mm, falling below the threshold before rising again to the threshold at a much higher altitude of approximately 200 Mm. Given our definition of critical height as the lowest altitude at which the decay index reaches the threshold ($n=1.5$), AR 11967 is assigned a relatively low critical height. This may explain the absence of a CME, as the associated flux rope likely reached the critical height but was prevented from erupting by the upper overlying field. This case highlights a potential limitation in our definition of critical height, which may not always accurately assess the eruptive potential of an AR. Therefore, caution must be exercised when interpreting critical height values derived from this definition.

\section{Conclusions} \label{sec:conclusion}
Our findings demonstrate that combining information on current neutralization with the critical height of the decay index provides a more comprehensive assessment of flux rope instability compared to using either metric individually. Assuming that all flaring activities in these ARs involved flux rope participation, we can apply torus instability analysis to provide a physical interpretation of our results. We proposed that the degree of current neutralization primarily reflects the net current within torus, thereby influencing the hoop force acting on the flux rope. Additionally, the critical height of decay index can serve as a metric of overlying magnetic field, which acts to counterbalance the hoop force. The critical height of the decay index, as presented in our study, provides crucial information regarding this strapping field. Therefore, by combining both, $|DC/RC|$ and the critical height of the decay index, we can comprehensively assess the torus instability conditions of a pre-existing flux rope within an active region.

\begin{figure}
\includegraphics[scale=0.5]{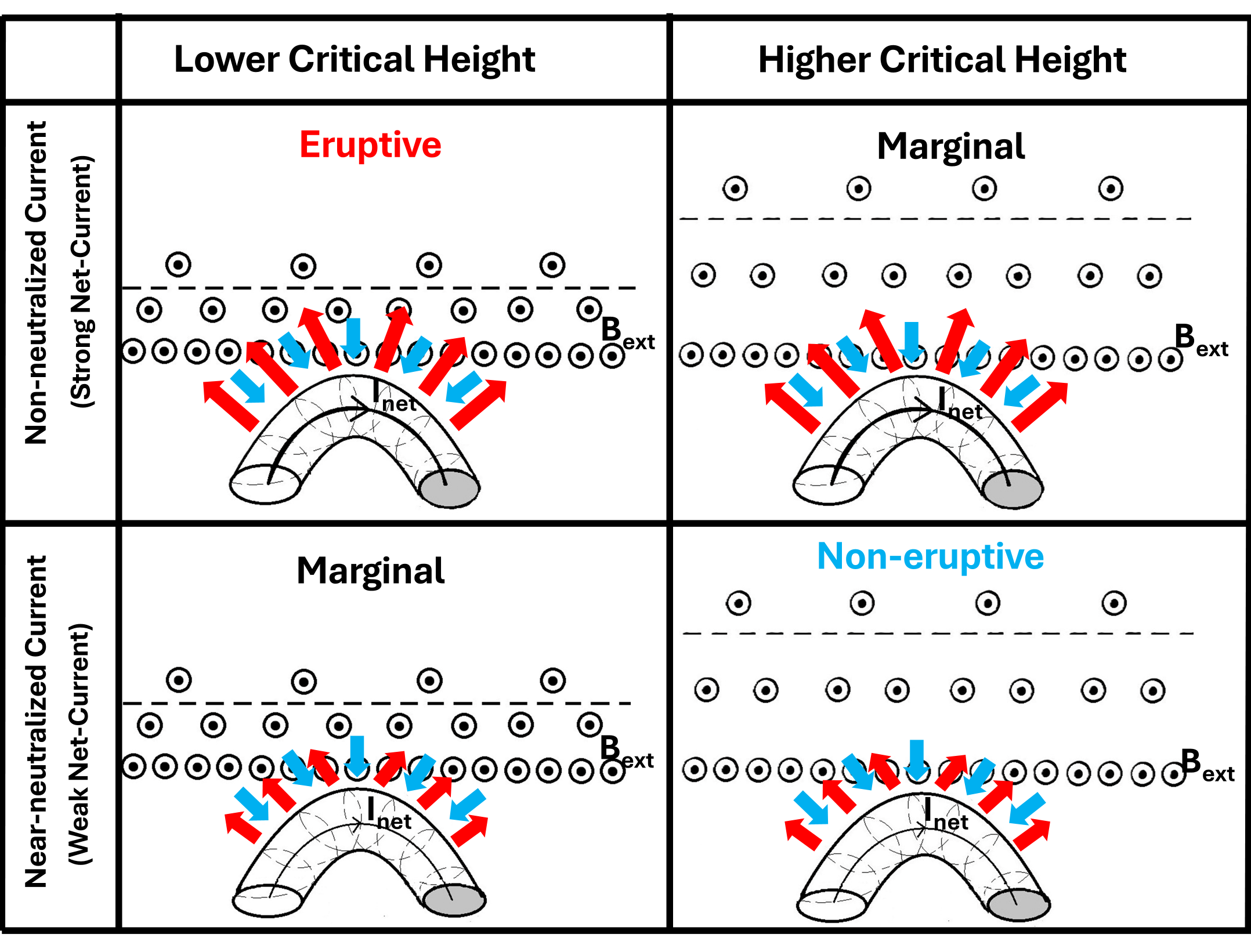}
\centering
\caption{Conceptual models of various possible conditions of magnetic field configurations, electric net-currents ($\boldsymbol{I_{net}}$), and forces acting on a flux rope in an AR. Red arrows directed outward the flux ropes represent the hoop force. Blue arrows directed toward the flux rope represent strapping forces  due to the presence of external magnetic fields ($\boldsymbol{B_{ext}}$), represented as circles with dots. Horizontal dashed line represents critical height of torus instability.}  \label{fig:cartoon}
\end{figure}

Summary of possible scenarios of degree of current neutralization and critical height conditions for a flux rope in an AR is shown in Figure \ref{fig:cartoon}. Here, near-neutralized current AR is represented as a weak net-current flux rope, while non-neutralized current AR is represented as flux rope with strong net-current.  Specifically, a flux rope characterized by a substantial net-current and a low critical height, exhibits a dominant hoop force, exceeding the constraining influence of the strapping field. Under these conditions, eruptive activity is highly probable. On the other hand, a flux rope possessing a limited net-current and positioned at an elevated critical height,  is less likely to erupt (non-eruptive). These two extreme scenarios offer relatively straightforward predictive outcomes. Moreover, intermediate conditions present greater complexity. A flux rope with a weak net-current situated within an AR of low critical height, or a flux rope with a strong net-current residing in an AR of high critical height, occupies a state of marginal stability. In our dataset, AR 11261 and 11967 can be considered to fall within these categories. In these situations, eruptivity greatly depends on the  significant perturbations in either the electric current ratio or the critical height. Such perturbations may arise from rapid emerging magnetic flux or horizontal photospheric motions. These intermediate cases pose a greater challenge for eruptivity prediction. Consequently, a more rigorous analysis on a small scale structure and photospheric evolution of the corresponding ARs is required to improve the prediction accuracy.           

We proposed a new non-dimensional parameter, S, which is the ratio of degree of current neutralization and critical height of decay index. We outline the methodology for deriving these parameters from magnetogram data and NLFFF models. Our findings indicate that eruptive flares tend to originate from ARs with significantly larger S values, exceeding unity. Conversely, non-eruptive flares are associated with ARs exhibiting smaller S values, closer to or below unity. These results suggest that eruptive flares occur in ARs with flux ropes characterized by strong hoop forces and low critical heights, rendering them susceptible to torus instability. In contrast, non-eruptive flares may arise from ARs with weaker hoop forces and high critical heights, insufficient to trigger torus instability. Consequently, the S-parameter offers a more effective means of distinguishing between eruptive and non-eruptive flares compared to using the degree of current neutralization or critical height alone.

While our results demonstrate strong alignment with torus instability analysis for most flare events, a few non-eruptive events exhibit deviation from these conditions. This discrepancy may arise from uncertainties in the selection of relevant magnetic flux, which can be influenced by limitations in NLFFF models or the pre-defined region of interest. Despite incorporating field line connectivity and flare brightenings into the selection process, the identification of relevant flux for current neutralization calculations remains somewhat subjective, potentially impacting the results. Additionally, the method used to determine the critical height may contribute to deviations. This indicates that our method for defining critical height is not without limitations and warrants further investigation. The maximum free-energy density point, while often suitable, may not always accurately represent the center of the flux rope. To improve decay index analysis, it may be better to consider the actual filament position, magnetic flux distribution of an AR, and 3D coronal geometry when selecting the location for evaluating the decay index profile. Despite this, we have shown that the proposed S-parameter have good potentials to discriminate between eruptive and non-eruptive flares. 

\begin{figure}
\includegraphics[scale=0.45]{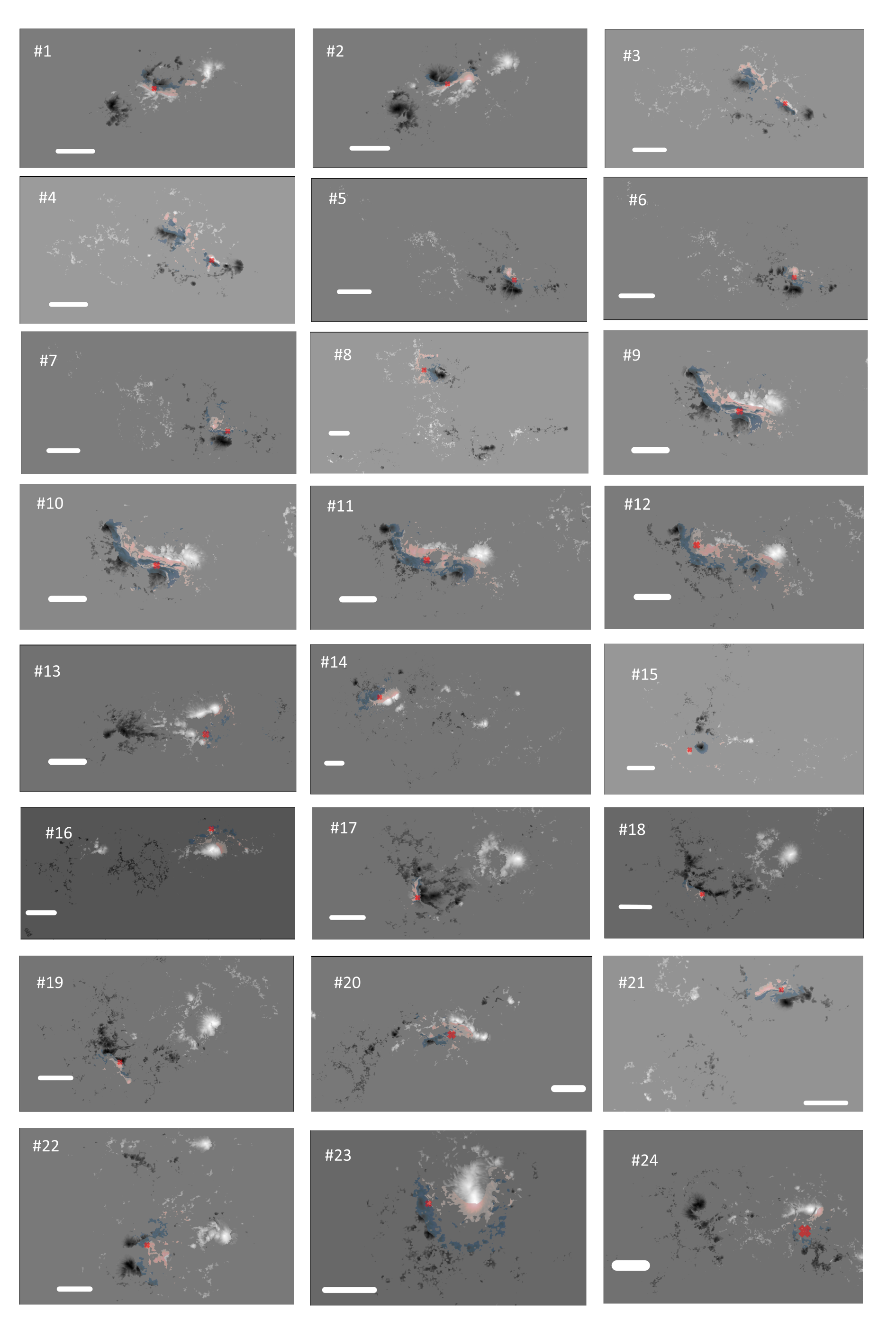}
\centering
\caption{$B_z$ of ARs from SHARP vector magnetograms with high confidence level of disambiguity for events \#1 to \#24. Selected magnetic fluxes and locations of maximum free-energy density points are shown as red (positive) and blue (negative) shaded regions, and red crossmarks respectively. Length of white solid lines represent scale of 40 Mm. \label{fig:fig10}}
\end{figure}

\begin{figure}
\includegraphics[scale=0.55]{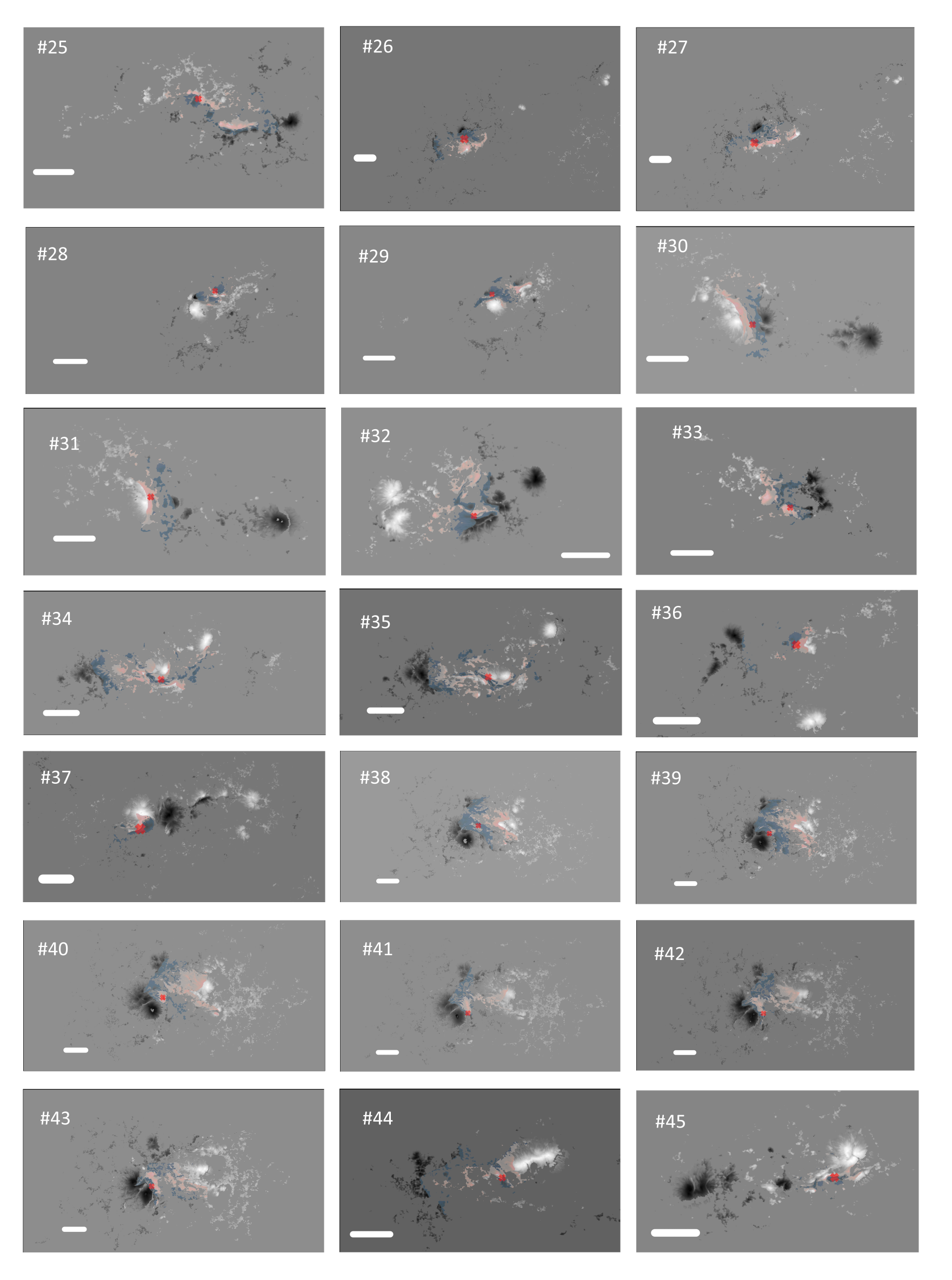}
\centering
\caption{$B_z$ of ARs from SHARP vector magnetograms with high confidence level of disambiguity for events \#25 to \#45. Selected magnetic fluxes and locations of maximum free-energy density points are shown as red (positive) and blue (negative) shaded regions, and red crossmarks respectively. Length of white solid lines represent scale of 40 Mm. \label{fig:fig11}}
\end{figure}

\textbf{Acknowledgements}
We thank the SDO/HMI team for producing vector magnetic field data products. JM acknowledges support by the National Research and Innovation Agency (BRIN). This work was carried out by the joint research program of Institute for Space–Earth Environmental Research, Nagoya University. This work was supported by JSPS KAKENHI Grant No. JP21H04492 (PI: K. Kusano). We acknowledge K.D. Leka for helpful discussions on the data analysis techniques used in this work.

\appendix
We present maps of $B_{z}$ of 45 ARs we used for our analysis in Figure \ref{fig:fig10} and \ref{fig:fig11}. Selected magnetic fluxes and locations of maximum free-energy density points are overlaid on $B_{z}$. 
\newpage

\bibliographystyle{aasjournal}
\bibliography{arxiv_paper_current_2024}{}





\end{document}